\documentclass[reprint,aps,prb,superscriptaddress]{revtex4-1}
\usepackage[utf8x]{inputenc}
\usepackage{amsmath}
\usepackage{graphicx}
\usepackage{amssymb}
\usepackage{txfonts}
\usepackage{color}

\newcommand{\dt}{\ensuremath{\partial_t}}
\newcommand{\dx}{\ensuremath{\partial_x}}
\newcommand{\fn}{\ensuremath{\mathbf{n}}}
\newcommand{\fni}{\ensuremath{\mathbf{n}^{(i)}}}
\newcommand{\fnj}{\ensuremath{\mathbf{n}^{(j)}}}
\newcommand{\heffi}{\ensuremath{\mathbf H^{(i)}_{\rm eff}}}
\newcommand{\jinter}{\ensuremath{J_{\rm inter}}}
\newcommand{\vsi}{\ensuremath{v_s^{(i)}}}
\newcommand{\vdw}{\ensuremath{v_{\rm DW}}}
\newcommand{\vdwe}{\ensuremath{v_{\rm DW1}}}
\newcommand{\vdwz}{\ensuremath{v_{\rm DW2}}}
\newcommand{\dvdw}{\ensuremath{\Delta v_{\rm DW}}}
\newcommand{\vpin}{\ensuremath{V_{\rm pin}^{(0)}}}
\newcommand{\kpara}{\ensuremath{K_{\parallel}}}

\begin{document}

\title{Dragging of magnetic domain walls by interlayer exchange}

\author{Martin Stier}
\author{Jennifer Erdmann}
\author{Michael Thorwart}
\affiliation{I. Institut f\"ur Theoretische Physik, Universit\"at Hamburg,
Jungiusstra{\ss}e 9, 20355 Hamburg, Germany}

\begin{abstract}
  Magnetic domain walls can be moved by spin-polarized currents due to spin-transfer torques. This opens the possibility to use them in spintronic memory devices as, e.g., in racetrack storage. Naturally, in miniaturized devices domain walls can get very close to each other and affect each others dynamics. In this work we consider two separated domain walls in different layers which interact via an interlayer exchange coupling. One of these walls is moved by a spin-polarized current. Depending on several parameters as the current density, the interlayer coupling or the pinning potential, the combined dynamics of the two domain walls can change very strongly allowing, e.g., for a correlated motion of the walls. In addition, more subtle effect appear as a suppression of the Walker breakdown accompanied by an increase of the domain wall velocity.
\end{abstract}

\maketitle   

\section{Introduction}

\begin{figure}[tb]
 \begin{minipage}[t]{.49\linewidth}
 (a) \includegraphics[width=\linewidth]{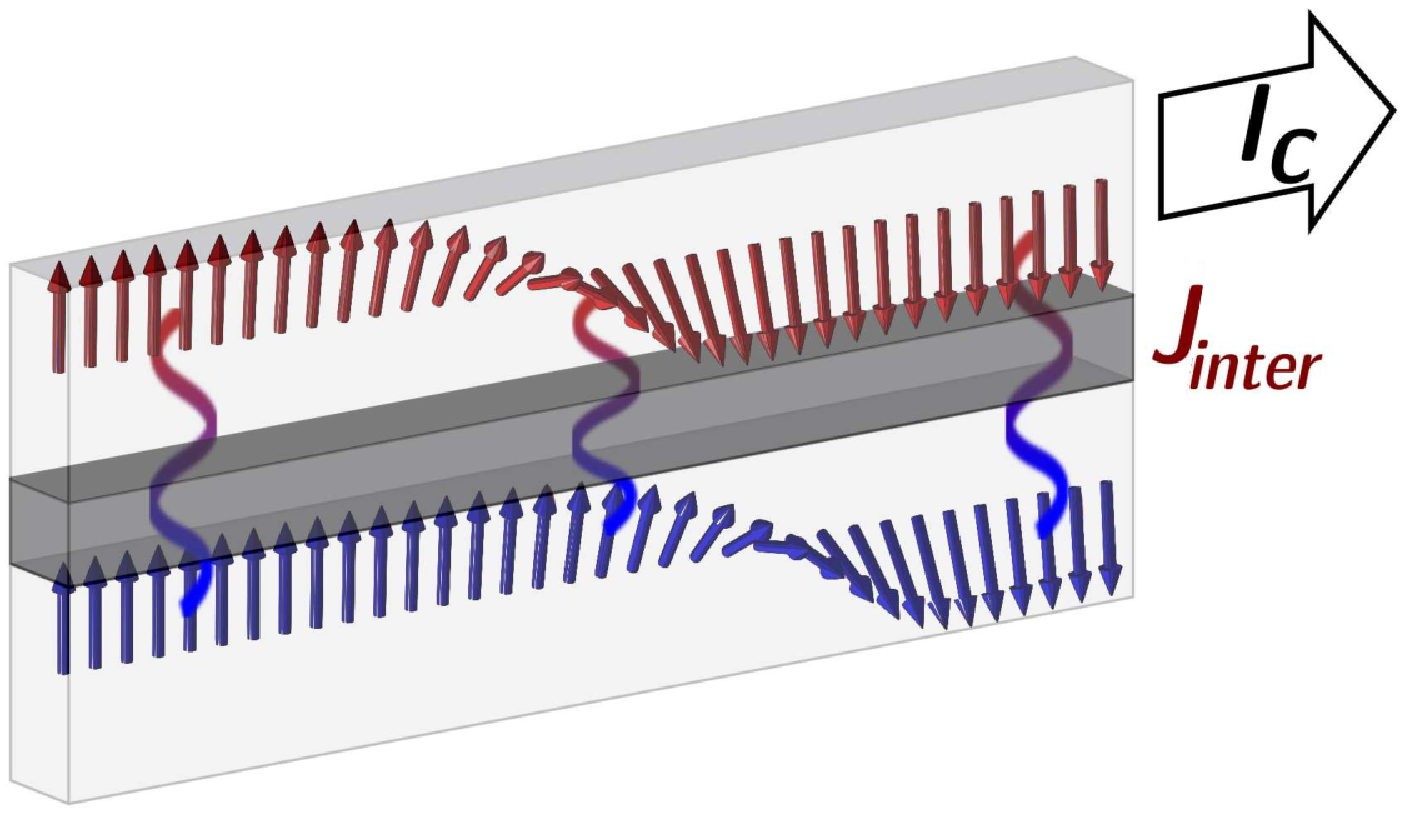}
 \end{minipage}
  \begin{minipage}[t]{.49\linewidth}
  (b)  \includegraphics[width=\linewidth]{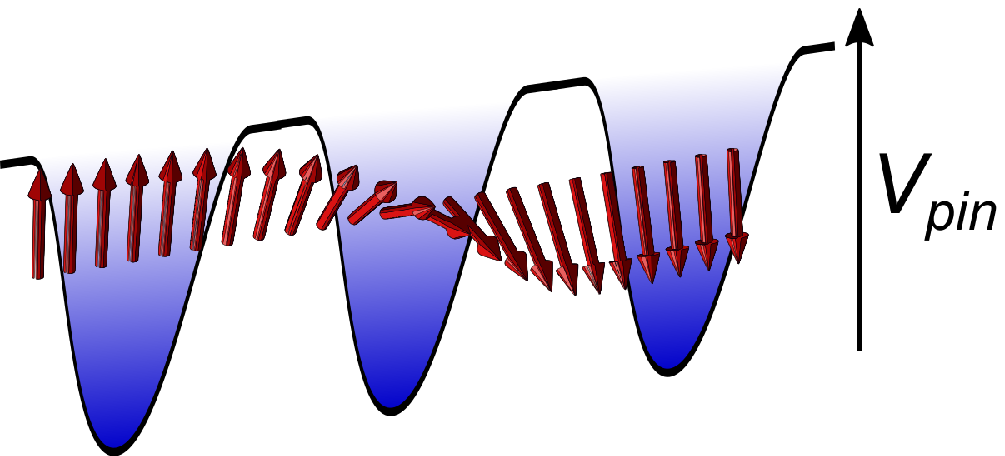}
  \end{minipage}
 \caption{\label{fig::scheme}
  (a) Two domain walls in separate layers but coupled via an IEC $\jinter$. One domain wall is moved by a spin-polarized current $I_c$ and may possibly drag the other wall due to the IEC. (b) Both DWs are affected by a periodic Gaussian pinning potential $V_{\rm pin}(x)$. The pinning centers for both DWs have the same shape and position.}
\end{figure}

The ongoing miniaturization of electronic devices increases the demand for high density information storage technology. Furthermore, in mobile devices this technology has to be very efficient in terms of energy consumption. A possible way to achieve this is to move magnetic domains on a racetrack memory \cite{parkin2008} by an electrical current instead of switching them by a magnetic field. Here, the electrons of a spin-polarized current generate a spin-transfer torque (STT) on the localized moments in the magnetic domains. This mechanism is well understood in terms of adiabatic \cite{berger1978} and non-adiabatic \cite{zhang2004} STTs. However, in presence of a spin-orbit interaction further torques may arise \cite{manchon_theory_sot,stier_non_eq_rashba} which may alter the dynamics of domain walls (DWs) \cite{risinggard2017,stier_chirality,wang2012,loconte_sot_driven_switching}. Furthermore, the effect of the spin texture on the STT itself has to be taken into account, particularly for small structures \cite{stier_nonlocal}.\\
Besides these effects, a more ordinary situation arises in the process of miniaturization. The lanes of a racetrack, and with it the magnetic domains, move closer to each other. Thus, these domains are more likely to interact, e.g., by the means of magnetostatic forces \cite{purnama_coupled_dws,obrien_near_field_ia,obrien_oscillations_of_coupled_dws,hayward_dw_imaging,lew_mirror_dws,thomas_dw_induced_coupling} or carrier mediated interlayer exchange coupling (IEC)\cite{hellwig_dws_in_af_coupled_films,Yang2015}. Even though these interactions may induce unwanted distortions from the single DW behavior, it has also been reported that coupled DWs suffer from the Walker breakdown (WB), a sudden drop of the velocity at a certain current threshold, only at higher current densities compared to single DWs \cite{Yang2015,purnama_coupled_dws}. Also the strength and the sign of the interlayer interaction can be tuned by the thickness of the spacer \cite{knepper_osc_interlayer_coupling,fert_review,Lavrijsen2012,stier_interlayer} allowing for tailoring of spin structures including artificial antiferromagnets \cite{Farokhipoor2014,duine2018synthetic}.\\
In this work we describe the motion of DWs coupled to each other by an interlayer exchange interaction as indicated in Fig. \ref{fig::scheme}. In contrast to Saarikoski et al. \cite{Tatara2014} only one of these DWs will be moved by a spin-polarized current while the second DW will only be able to react to the change of the position of the first wall via the IEC. We will present systematic investigations of the dynamics of these two walls in a large parameter space including current density, IEC strength and a pinning potential. Our results generally confirm a possible drag of the second DW as well as a possible suppression of the WB. 

\section{Model}

For the sake of simplicity we restrict the system to two one-dimensional layers each containing a ferromagnetic DW described by the magnetic texture $\mathbf M^{(i)}(x,t) = M_s \fni (x,t)$ with the layer index $i =1,2$. The dynamics of the according textures are calculated by coupled Landau-Lifshitz-Gilbert equations (LLGs)
\begin{align}
 \dt \fni =& - \gamma\fni \times \heffi + \alpha \fni \times \dt\fni\nonumber \\
 &+ \vsi (1- \beta \fni\times) \dx \fni \label{eq::llg}\\
 &-\frac{\jinter}{\hbar} \fni\times \fnj\nonumber, \quad i\neq j
\end{align}
In this expression $\mathbf{H_\text{eff}}$ is the effective field, $\alpha$ the Gilbert damping parameter, $\beta$ the non-adiabatic spin transfer torque parameter and $J_{\text{inter}}$ the IEC strength. The spin velocity is defined by $v_s=\frac{a^3\mathcal{P}I_c}{2S|e|}$, in which $\mathcal{P}$ is the current polarization, $a$ the lattice constant, $S$ the spin density and $e$ the elementary electric charge. The first term of the LLG describes the adiabatic precession of the spin around the magnetic field, whereas the second term depicts the damping, the third term includes the adiabatic and non-adiabatic spin transfer torque \cite{zhang2004}, which results from the spin polarized current. Finally, the influence of the IEC is characterized by the forth term where each magnetization feels the magnetization of the other DW as an effective field.

\section{Results}

In this work, we only treat magnetic Bloch DWs with the initial magnetization profile
\begin{equation}
 \fni(x) = {\rm cosh}^{-1}\left(\frac{x-x_0^{(i)}}{\lambda_{\rm DW}}\right)\mathbf{\hat y}+{\rm tanh}\left(\frac{x-x_0^{(i)}}{\lambda_{\rm DW}}\right)\mathbf{\hat z}\ .\label{eq::mag}
\end{equation}
Both DWs are coupled ferromagnetically to each other which is, in the classical picture, completely analogous to the antiferromagnetic case when we replace $\jinter\to -\jinter$ and $\fn^{(2)}(x)\to -\fn^{(2)}(x)$. We fix the parameters such that $\hbar\gamma H_{\rm eff} = J \partial^2_x\fn +\kpara n_z - K_{\perp} n_x$ with $\lambda_{\rm DW}=\sqrt{J/\kpara}=5$nm and $K_{\perp} = \kpara/2$. Furthermore, we set the Gilbert damping $\alpha=0.1$, the polarization of the spin-polarized current $\mathcal P = 1$ and the lattice constant $a=0.5$nm that $v_s =100{\rm m/s} \Leftrightarrow I_c = 1.3\times 10^{11}{\rm A/m^2}$. All results are achieved by solving the LLG numerically by Runge-Kutta techniques.

\begin{figure}[tb]
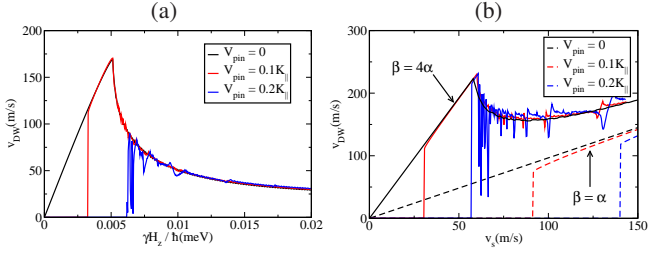

 \begin{minipage}[t]{.49\linewidth}
  (a)\\ \includegraphics[width=\linewidth]{v_vs_b2}
 \end{minipage}
  \begin{minipage}[t]{.49\linewidth}
  (b)\\ \includegraphics[width=\linewidth]{v_vs_vs}
  \end{minipage}
 \caption{\label{fig::vDW}Domain wall velocity vs (a) external magnetic field $H_z$ or (b) applied current $v_s$ for different pinning potential strengths. For finite pinning a critical field/current has to be overcome to move the DW. Above the Walker breakdown, either field- or current-induced, the dependence of the velocity on the field/current gets noisy.}
\end{figure}

\subsection{Current-induced vs field-induced motion}

The current-induced dynamics of coupled DWs is basically a complex combined current- and field-induced motion. Many aspects of this combined dynamics may already be understood from the single dynamics which we show in Fig. \ref{fig::vDW}. It is crucial to keep in mind that the DW's velocity is in general not necessarily proportional to either the magnetic field or the electric current. In particular, in presence of a hard magnetic axis the field-induced motion experiences a WB resulting in a peak of the DW velocity and thus in a decrease of $\vdw$ at higher fields. As we will find, this limits the maximum effective coupling between two DWs coupled by a Heisenberg-exchange $\propto\jinter$. Similarly, a current-induced WB also appears when the non-adiabaticity parameter $\beta$ exceeds the Gilbert damping $\alpha$. The occurrence of this WB is are strong limitation of the DW velocities achievable by current-induced motion. A prevention of the WB is desirable for an efficient application of DWs in possible devices. These generic features of DW motion are well understood and extensively discussed in, e.g., Refs. \cite{lucassenphd,mougin2007}.

\subsection{Pinning}

Domain walls under the influence of neither an external field nor an applied electrical current do not move even in absence of a pinning potential within a wire. When two DWs are coupled, however, they either attract or repel each other depending on the sign of $\jinter$. In this work we will only focus on the case of attraction. Thus even for low interactions, two DWs will end at the same position when we simply wait long enough after we switch off the external field or the current. To go beyond these predictable results we introduce an effective pinning potential which changes the easy axis anisotropy to
\begin{equation}
 K_{\parallel}\to K'_{\parallel}= K_{\parallel} - V_{\rm pin}(x)~.
\end{equation}
Of course, pinning will be present in real materials anyway. For the sake of simplicity we choose a periodic Gaussian potential, see Fig. \ref{fig::scheme},
\begin{equation}
V_{\rm pin}(x)= V_{\rm pin}^{(0)} \sum_r \exp \left[-(x-r\Delta x_{\rm pin})^2/w_{\rm pin}^2\right]\ ,
\end{equation}
where each Gaussian has a width of $w_{\rm pin}$ and a distance $\Delta x_{\rm pin}$ to the next one. Throughout the work, we set $w_{\rm pin}=\lambda_{\rm DW}$ and $\Delta x_{\rm pin}=5\lambda_{\rm DW}$, where $\lambda_{\rm DW}$ is the DW width. The pinning potentials of the two coupled DWs are identical to each other.\\
The effect of the pinning potential on the dynamics of the uncoupled DWs can be seen in Fig. \ref{fig::vDW}. We find that a critical field, or current respectively, is needed to move the DWs. As expected, the necessary critical field/current increases with increasing pinning strength $\vpin$. Note that in the regime of the WB the depinning of the DW gets more unpredictable, i.e., the DW velocities fluctuate strongly. This happens due to the tilting of the DW for $\alpha\neq \beta$ which allows for an easier depinning\cite{Parkin190}.

\begin{figure}[tb]
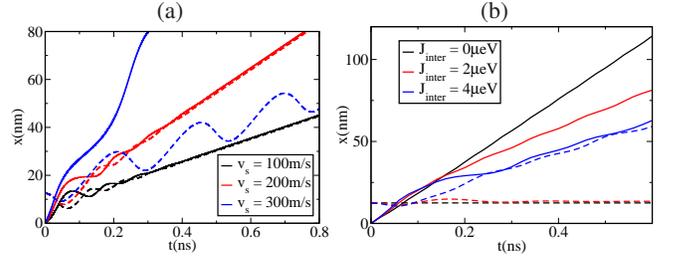

 \centering
 \begin{minipage}[t]{.49\linewidth}
 (a) \includegraphics[width=\linewidth]{x_vs_t2}
 \end{minipage}
  \begin{minipage}[t]{.49\linewidth}
  (b) \includegraphics[width=\linewidth]{x_vs_t}
  \end{minipage}
  \caption{\label{fig::x_vs_t} Time-dependent position of coupled DWs. Solid lines show the position of the current-driven DW and dashed lines that of the dragged DW for (a) vanishing pinning potential or for (b) finite pinning potential. Without pinning, both DWs move at the same velocity as long as the current-induced velocity of the first DW does not exceed the maximum field-induced velocity. When pinning is present, a critical coupling has to be overcome to move the second DW. Parameters are $\alpha=\beta=0.1$ and (a) $\jinter= 10\mu\rm eV$, $\vpin=0$ or (b) $v_s=150\rm m/s$, $\vpin=0.2K_{\parallel}$.
}
\end{figure}

\subsection{Coupled motion}
\begin{figure}[tb]
  \centering
\includegraphics[width=.8\linewidth]{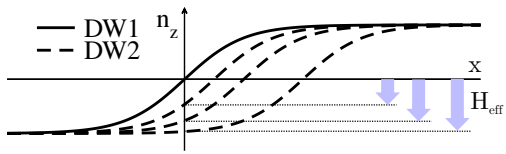}
\caption{\label{fig::eff_field}Magnetization profile $n^{(1,2)}_z(x)$ of two Bloch DWs. The according effective field in $z$ direction $H^{\rm eff,z}_{\rm inter} = \jinter n_z^{(2)}(x\approx 0)$ induced on the first DW by the interlayer coupling depends on the distance of the DWs. Thus, the DWs can adjust the field to an optimal distance to match their velocities in certain limits.}
\end{figure}
Basically, the coupled motion of two DWs is a combination of a field- and a current-induced motion even though in a rather complex and dynamic way as the external field, i.e., the field created by the IEC, is non-constant in time and space. Despite this more complex behavior, we can still explain the principles of the combined motion in most aspects by the dependencies of the DW velocity shown in Fig. \ref{fig::vDW}. Initially, the first DW has to be moved by the applied current following essentially the dynamics of the purely current induced motion. Then, it can try to drag the second DW by means of a field-induced motion. Basically, three scenarios may arise: (i) the applied current is not sufficient to loosen the first DW from the pinning potential, (ii) the current is sufficient to loosen the first DW but the IEC coupling is not strong enough to move the second DW, and (iii) both the current and the coupling are strong enough to move both DWs. The first scenario is trivial, as it is essentially static. Thus, we will not discuss it in detail. The latter two scenarios are shown in Fig. \ref{fig::x_vs_t}. To show the dragging effect we let the first, current-driven, DW start in an position behind the second, potentially dragged, DW. Forgetting about pinning for the moment, we find that both DWs will move with the same velocity in a relatively broad current regime, as can be seen in Fig. \ref{fig::x_vs_t}. This is possible as the DWs can adjust their effective fields induced on each other by changing their distance. When they come closer, they reduce these fields which we show in Fig. \ref{fig::eff_field}. When the maximum effective field $H^{\rm inter,max}_{\rm eff} = |\jinter|$ for $n_z = 1$ is very strong, i.e., has a value on the right side of the peak in the $\vdw$--$H$ diagram [cf. Fig. \ref{fig::vDW} (a)], a reduction of distance, and with it, of $H^{\rm inter}_{\rm eff}$ yields an increase of the velocity and the DWs can match their speed. Obviously, this is only possible if the velocity of the current-driven DW does not exceed the maximum speed achievable by field-induced motion. In that case the field-driven DW cannot hold up with the current-driven DW and will lag behind.\\
We now add pinning to the coupled motion where the pinning potential, and in particular, the positions of pinning centers, are the same for both DWs. Again, the first DW will start from a position behind the second DW and both DWs are in a local minimum of the pinning potential. Thus, without an applied current or an IEC, they rest at their according positions. In Figure \ref{fig::x_vs_t}, we have chosen a current which is strong enough to free the first DW from its pinning center. However, only with a sufficient IEC it is possible to also move the other DW. Obviously, the coupled motion of the DWs is strongly parameter dependent. We will discuss a comprehensive picture in the next section.

\section{Parameter dependencies of coupled domain wall motion}

\begin{figure*}[tbp]
 \includegraphics[width=\linewidth]{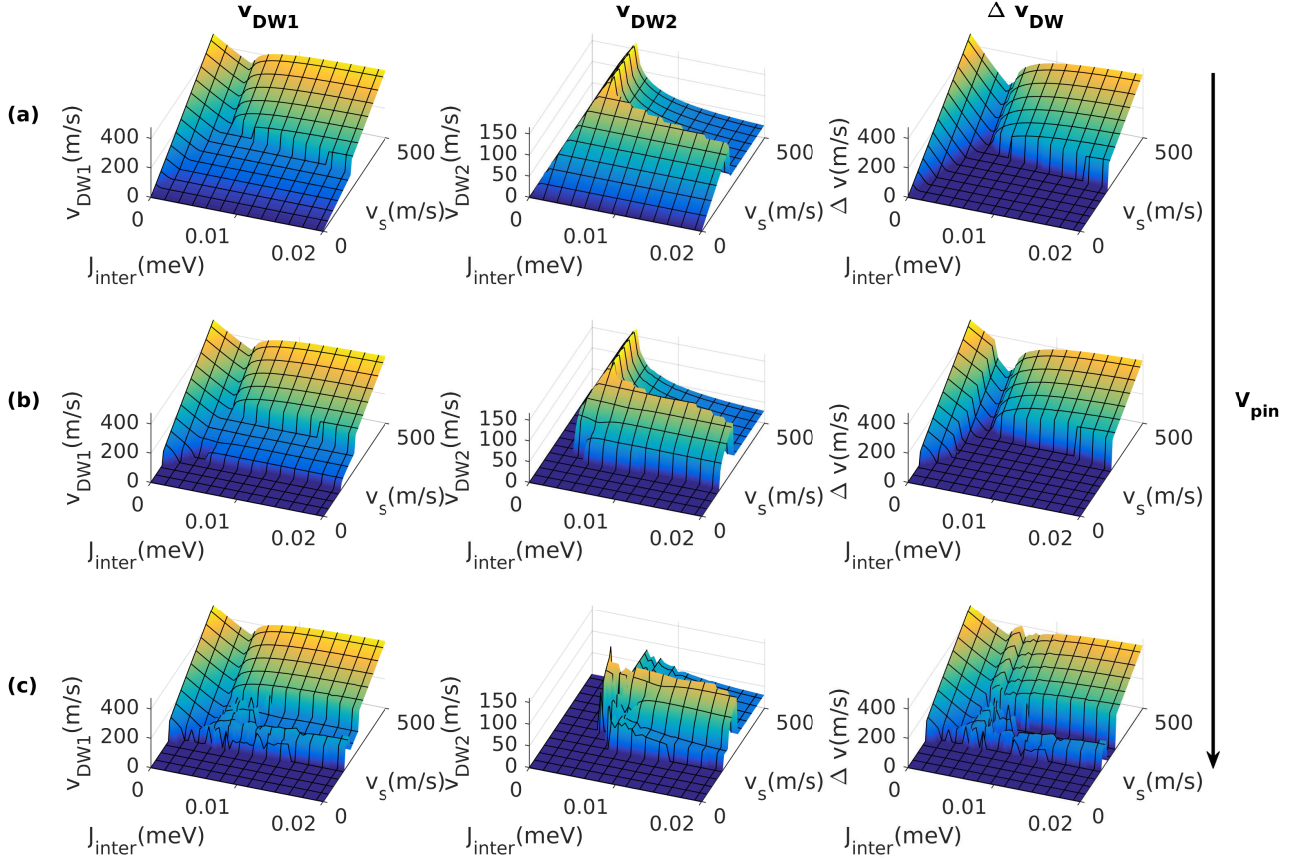}
 \caption{\label{fig::Jinter_vs_vs}
  Velocity of first (current driven) DW $v_{\rm DW1}$, second (dragged) DW $v_{\rm DW2}$ and velocity difference $\Delta v = [x_{\rm DW1}(t) -x_{\rm DW2}(t)]/t$ vs IEC $\jinter$ and spin-polarized current $v_s$ at different pinning potential strengths: first row (a) $V_{\rm imp} = 0$, second row (b) $V_{\rm imp} = 0.1K_{\parallel}$ and third row (c) $V_{\rm imp} = 0.2K_{\parallel}$. All: $\beta=\alpha$. For low velocities of the first DW, the second DW is able to follow the first DW due to the field-induced drag. At higher velocities $v_{\rm DW1}$, the second DW reaches its maximum field-induced velocity. This maximum velocity cannot be enhanced by $\jinter$ as a field-induced WB occurs around $\jinter=0.005$meV. Finite pinning potentials require sufficiently strong spin-polarized currents $v_s$ to move the first wall and additionally sufficiently large couplings $\jinter$ to move the second DW.}
\end{figure*}

In this section we will show how the coupled motion of two DWs depend on the three most important parameters of our model, $\jinter$, $v_s$ and $\vpin$. The dynamics of the DWs will be discussed by their respective long term averaged velocities $\vdwe$ and $\vdwz$ as well as by the velocity difference $\Delta v$. In Fig. \ref{fig::Jinter_vs_vs}, we show the dependence of these velocities on $\jinter$ and $v_s$. We have put $\alpha=\beta$ in this case, which means the current driven DW experiences no WB by itself. Thus, in the unpinned and uncoupled case, the $\vdwe(v_s)$ curve is just a straight line (cf Fig. \ref{fig::Jinter_vs_vs}a at $\jinter=0$). This makes it easier to see the effect of the coupling. For $\vpin = 0$, the purely interlayer exchange field-driven DW follows the generic curve of Fig. \ref{fig::vDW}a almost exactly. At low currents $v_s$, it can keep up with the speed of the first DW almost everywhere. The higher the current, the faster the first DW and the second DW can only follow at its peak velocity around $\jinter\approx 0.005$meV. As soon as the maximum velocity is exceeded by the first wall, the second DW is left behind which increases the effective field (cf. Fig. \ref{fig::eff_field}) and thus decreases the second DWs velocity. We find from regions where $\Delta v_{\rm DW}=0$ that both DWs move together, up to a current of $v_s\approx 200$m/s and slightly higher for optimal values of $\jinter$. This broad parameter regime is only made possible by the self-adjustment of the effective interlayer exchange field explained in Fig. \ref{fig::eff_field}.\\
The introduction of pinning to the system makes the situation more complex as a finite critical $v_s$ is needed to move the first DW as well as a critical $\jinter$ to also move the second DW. Even though, the region of $\dvdw = 0$ seems to be hardly changed for $\vpin = 0.1\kpara$ (cf. Fig. \ref{fig::Jinter_vs_vs}b) it does not mean that both DWs move together in the whole parameter region. For approximately $v_s < 100$m/s, which covers half of the $\dvdw=0$ region, both DWs do not move at all. Naturally, the second DW moves in an even smaller region, as it needs the first DW to move at all and a sufficient $\jinter$. When both $v_s$ and $\jinter$ exceed their respective critical value, both DWs move and the picture resembles the unpinned case. A stronger pinning (Fig. \ref{fig::Jinter_vs_vs}c) yields a slightly more unpredictable dynamics of the DWs in the vicinity of the depinning region which becomes manifest in fluctuations of the DW velocities.\\  
\begin{figure*}[tbp]
 \includegraphics[width=\linewidth]{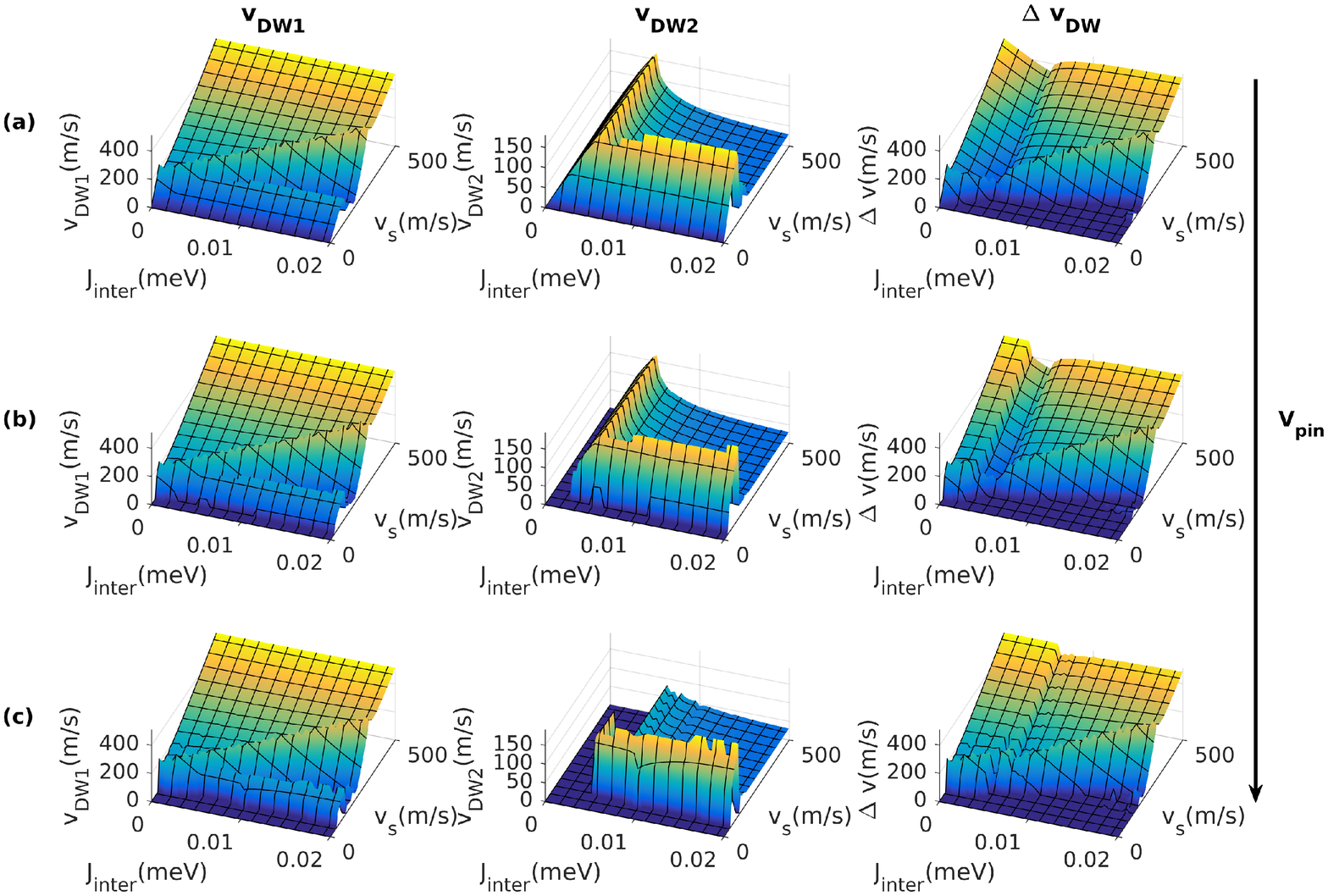}
 \caption{\label{fig::Jinter_vs_vs_b=4a}
  Same as Fig. \ref{fig::Jinter_vs_vs} but with $\beta=4\alpha$. This allows for an easier depinning of the first DW. In addition, a current induced WB occurs for the first DW, with slight impact on the second DW. Due to the IEC, the current induced WB is shifted to higher currents.}
\end{figure*}
When we set $\beta \neq \alpha$, as we have done in Fig. \ref{fig::Jinter_vs_vs_b=4a} by choosing $\beta=4\alpha$, we introduce an additional feature -- the current-induced WB. We now find an additional peak in the DW velocity according to its $v_s$-dependence (also cf. Fig. \ref{fig::vDW}). The position of this peak in the uncoupled case is determined by the anisotropies \cite{mougin2007}. However, we find a shift of this peak to higher values of $v_s$ with increasing $\jinter$. This is because, in addition to the simplified picture drawn in Fig. \ref{fig::eff_field}, the DWs do not only induce a field in $z$ direction on each other but also contain the $x,y$ directions [cf. also Eq. (\ref{eq::mag})]. These act as effective anisotropies and stabilize the DW motion, i.e., they prevent a precessional motion which marks the onset of the WB. Due to this stabilization both, or at least the first DW, can move faster than an uncoupled DW. This prevention of the current induced WB has also be found in artificial antiferromagnets\cite{Tatara2014,duine2018synthetic,Gomonay2017}. Furthermore, we find, in agreement with Fig. \ref{fig::vDW}, that DWs are depinned more easily for $\alpha<\beta$. Thus, the dynamics of coupled DWs in terms of fast movement benefits from a strong non-adiabaticity $\beta$ overall.\\
\begin{figure*}[tbp]
 \includegraphics[width=\linewidth]{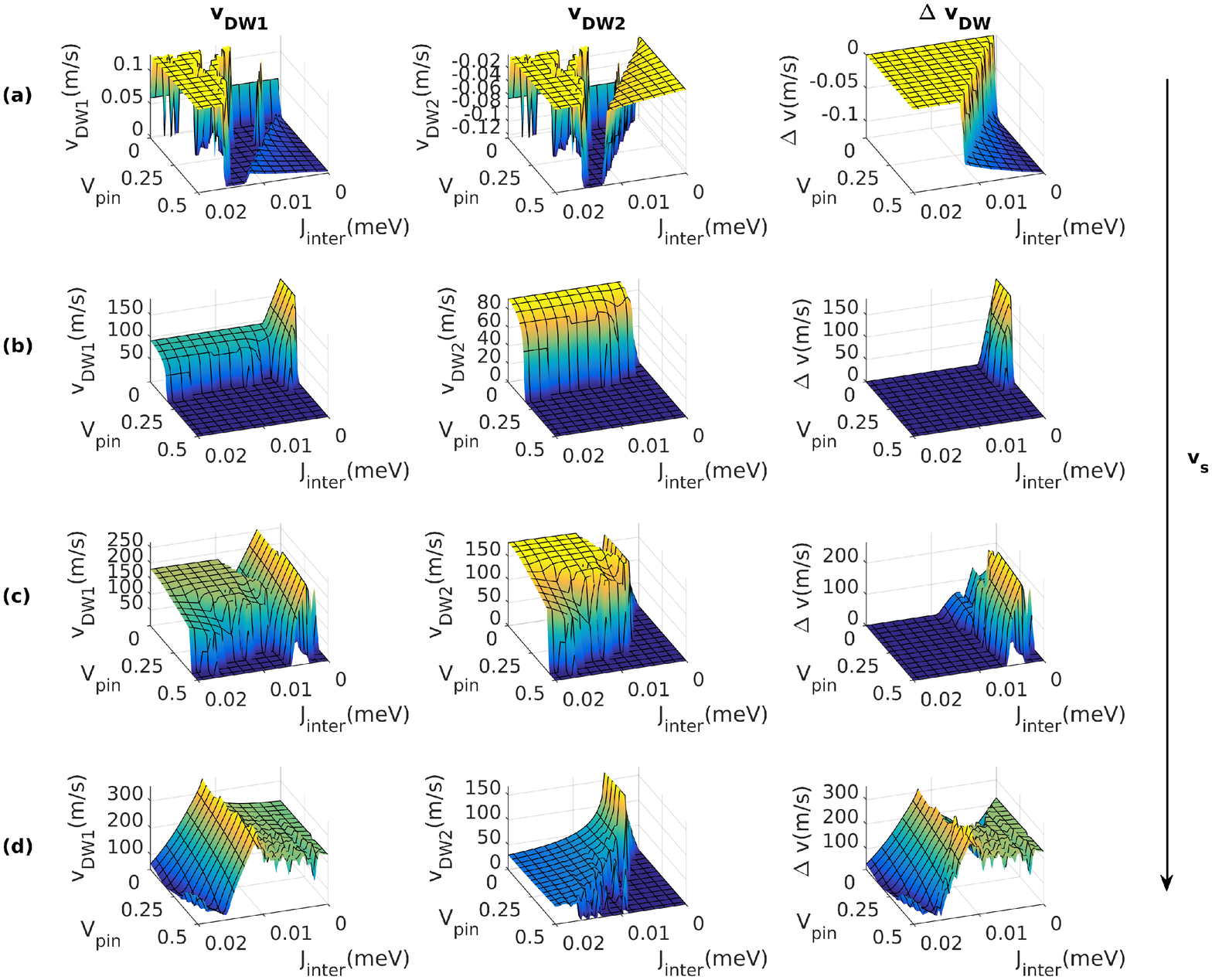}
 \caption{\label{fig::Jinter_vs_Vimp_b=4a}
 Velocity of first (current driven) DW $v_{\rm DW1}$, second (dragged) DW $v_{\rm DW2}$  and velocity difference $\Delta v = [x_{\rm DW1}(t) -x_{\rm DW2}(t)]/t$ vs IEC $\jinter$ and pinning potential strength $V_{\rm imp}$ at different spin-polarized currents: first row (a) $v_s = 0$, second row (b) $v_s=50\rm m/s$, third row (c) $v_s=100\rm m/s$ and forth row (d) $v_s=200\rm m/s$. All: $\beta=4\alpha$. Even for zero current, the DWs move slowly towards each other and may overcome the pinning potentials for sufficient IEC and end up at the same positions [cf. row (a)]. For small $\jinter$, the first DW is depinned more easily as is does not have to drag the second DW. The pinning gets less relevant for higher current densities $v_s$.}
\end{figure*}
For a systematic view on the necessary coupling between the DWs to depin the DWs from their pinning potential, we show Fig. \ref{fig::Jinter_vs_Vimp_b=4a}. From the current-free case (Fig. \ref{fig::Jinter_vs_Vimp_b=4a}a) we find from the $\Delta v_{\rm DW} = [x_{\rm DW1}(t_{\rm end}) - x_{\rm DW2}(t_{\rm end}) ]/t_{\rm end}\neq 0$ the region where $\jinter$ is not sufficient to move the DWs towards each other, i.e. to depin them. As the second DW is only moved by the interlayer exchange, we would naively expect this region to be the same for all currents flowing through the first DW. This turns out to be wrong as, at least for $\alpha\neq\beta$, the first DW is tilted\cite{stier_chirality} by the non-adiabatic STT $\propto(\beta-\alpha) v_s$ and thus the field exerted on the second DW is changed. In fact, there are finite velocities $\vdwz$ even in regions where the DWs could not be depinned in the current-free system. 
\begin{figure*}[tbp]
 \includegraphics[width=\linewidth]{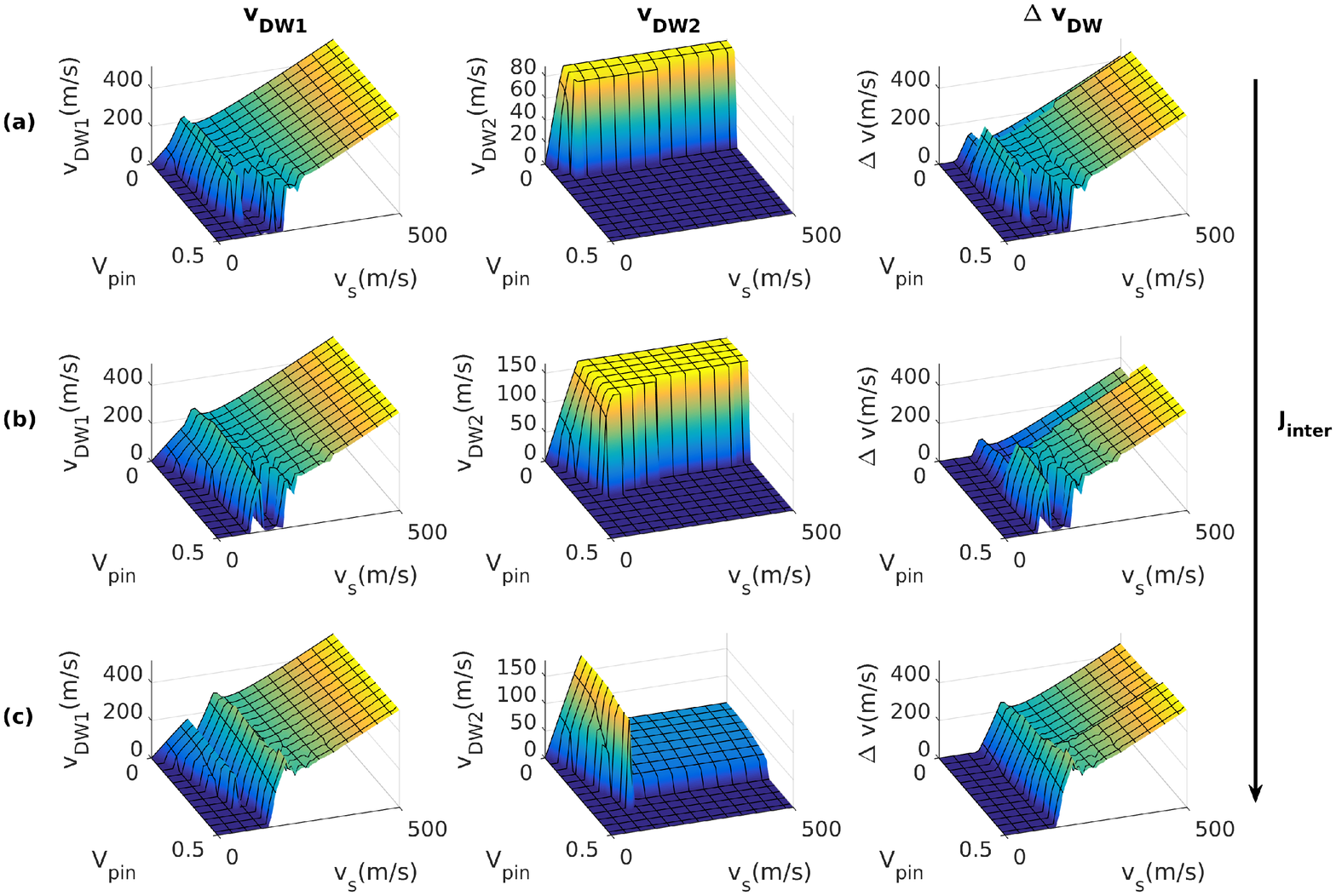}
 \caption{\label{fig::vs_vs_Vimp_b=4a}
  Velocity of first (current driven) DW $v_{\rm DW1}$, second (dragged) DW $v_{\rm DW2}$  and velocity difference $\Delta v = [x_{\rm DW1}(t) -x_{\rm DW2}(t)]/t$ vs pinning potential strength $V_{\rm imp}$ and spin-polarized current $v_s$at different IEC: first row (a) $\jinter = 0.0025$meV, second row (b) $\jinter = 0.005$meV and third row (c) $\jinter = 0.01$meV. All: $\beta=4\alpha$. Note that the current $v_s$ indirectly influences the depinning of the second DW at least slightly.}
\end{figure*}
We can see this even better in Fig. \ref{fig::vs_vs_Vimp_b=4a} where we plot the DW velocities vs. $\vpin$ and $\jinter$. Here, $v_s$ slightly changes the region $\vdwz>0$ even for large $v_s$. In this case, the DWs are only close to each other over a very short time. Only when the first DW passes the second one, the tilt of the first DW can be felt by the second DW and can have an effect on its depinning.\\
All of these effects have to be considered in technical applications, particularly when DWs are close to each other, for example in systems with many DWs. Without pinning, an independent motion of DWs, and with it a reasonable manipulation of stored data, would be impossible even for very tiny IECs. Obviously, a pinning will always be present in real materials and as such will act beneficial in terms of stabilizing the stored information. The IEC will still provide a limiting factor in decreasing the distance of neighboring wires. On the other hand, on can think of several positive effects of the IEC. First, there is the suppression of the WB which allows for faster DW motion. But one could also use dedicated DWs in extra layers to provide movable distinguished positions in a wire, similar to modifiable anisotropies.

\section{Summary}

In this work we investigated the dynamics of to coupled magnetic domain walls systematically. The DWs are coupled via an interlayer exchange which we model by a Heisenberg term. Furthermore, both DWs feel the presence of a pinning potential. Since only one DW is driven by an electric current, the other one can only be moved by the interlayer exchange. By means of an extended Landau-Lifshitz-Gilbert equation we calculated the DWs' velocities in dependence of important parameters. We find that a dragging of the second DW is possible in a rather broad parameter range in particular for an effective interlayer exchange which is not to be confused with a strong interlayer exchange. Beside the dragging of one DW, we find the phenomenon of a suppressed Walker breakdown. Due to this, both DWs can move faster than uncoupled current-driven DWs in a certain parameter regime. This results of this work show how to possibly optimize DW dynamics in very dense, coupled DW systems.

\begin{acknowledgments}
We acknowledge support from the Deutsche Forschungsgemeinschaft (DFG).
\end{acknowledgments}

\providecommand{\WileyBibTextsc}{}
\let\textsc\WileyBibTextsc
\providecommand{\othercit}{}
\providecommand{\jr}[1]{#1}
\providecommand{\etal}{~et~al.}

\clearpage

\end{document}